\documentclass{article}

\usepackage{PRIMEarxiv}

\usepackage{graphicx}
\usepackage{epstopdf, epsfig}

\usepackage{amsmath}
\usepackage{amssymb}
\usepackage{mathtools}
\newcommand{\N}{\mathbb{N}}
\newcommand{\defeq}{\coloneqq}
\newcommand{\upi}{\mathrm{i}}

\usepackage{url}
\usepackage{natbib}

\usepackage[utf8]{inputenc} 
\usepackage[T1]{fontenc}    

\title{A mechanistic evaluation of the local Bloch wave approximation in graded arrays of vertical barriers}

\author{
  Ben Wilks \\
Department of Mathematics and Statistics\\
University of Otago\\
PO Box 56\\
Dunedin 9054\\
New Zealand\\
  \texttt{wilbe612@student.otago.ac.nz} \\
   \And
 Fabien Montiel \\
Department of Mathematics and Statistics\\
University of Otago\\
PO Box 56\\
Dunedin 9054\\
New Zealand\\
  \texttt{fmontiel@maths.otago.ac.nz} \\
  \And
 Sarah Wakes \\
Department of Mathematics and Statistics\\
University of Otago\\
PO Box 56\\
Dunedin 9054\\
New Zealand\\
  \texttt{sarah.wakes@otago.ac.nz}
}

\begin{document}
\maketitle

\begin{abstract}
Wave interaction with graded metamaterials exhibits the phenomenon of rainbow reflection, in which broadband wave signals slow down and separate into their frequency components before being reflected. This phenomenon has been qualitatively understood by describing the wave field in the metamaterial using the local Bloch wave approximation (LBWA), which locally represents the wave field as a superposition of propagating wave solutions in the cognate infinite periodic media (so-called Bloch waves). We evaluate the performance of the LBWA quantitatively in the context of two-dimensional linear water-wave scattering by graded arrays of surface-piercing vertical barriers. To do this, we implement the LBWA numerically so that the Bloch waves in one region of the graded array are coupled to Bloch waves in adjacent regions. This coupling is computed by solving the scattering of Bloch waves across the interface between two semi-infinite arrays of vertical barriers, where the barriers in each semi-infinite array can have different submergence depths. Our results suggest that the LBWA accurately predicts the free surface amplitude across a wide range of frequencies, except those just above the cutoff frequencies associated with each of the vertical barriers in the array. This highlights the importance of decaying Bloch modes above the cutoff in rainbow reflection.
\end{abstract}

\section{Introduction}
Graded metamaterials, i.e. media composed of a spatially-varying resonant substructure, have been shown to exhibit remarkable properties across a wide range of wave sciences, including electromagnetic waves \citep{Tsakmakidis2007}, acoustic waves \citep{Zhu2013,Ni2014,Jimenez2017,Bennetts2019,Zhao2019}, elastic waves \citep{Skelton2018,Arreola-Lucas2019}, seismic waves \citep{Colombi2016} and water waves; the latter area is the focus of this paper. Modelling has shown that water wave metamaterials can locally amplify wave energy; the subsequent absorption or reflection of this energy may have important applications in the design of wave energy converters and coastal breakwaters, respectively \citep{Bennetts2018,wilks2022rainbow}. Grading the geometric or physical properties of the resonators in the metamaterial can result in broadband effects.

The most commonly used conceptual tool for understanding wave propagation through graded metamaterials has been the \textit{local Bloch wave approximation} (LBWA). This approach assumes that if the grading of the metamaterial is sufficiently weak, then the wave field inside the metamaterial can locally be approximated as a sum of propagating solutions of the Bloch-Floquet problem, in which the unit cell of the infinite array matches the local geometry of the metamaterial. These propagating solutions, which are associated with a frequency-dependent group velocity, only exist in frequency intervals known as passbands. In the complementary frequency intervals (known as stopbands), waves cannot propagate, i.e. only decaying modes are present. The LBWA explains the properties of graded metamaterials as follows:
\begin{itemize}
    \item Waves of a given frequency enter the metamaterial and gradually slow down because the group velocity of the local Bloch wave is decreasing.
    \item When the group velocity becomes zero at the so-called \textit{turning point}, propagation further into the metamaterial is forbidden because the local geometry only supports decaying modes.
    \item Since propagation beyond the turning point is impossible, the local forward-propagating wave must couple with the local backward-propagating wave at the turning point, which continues to propagate backwards through the array, ultimately leading to reflection.
    \item The near-zero group velocity of the forward-propagating and backward-propagating Bloch wave near the turning point results in local energy amplification.
    \item The grading causes the location of the turning point to be frequency dependent, therefore allowing the metamaterial to spatially separate broadband signals into their frequency components.
\end{itemize}
Variations of the above argument can be found in a large number of research papers on the topic of graded metamaterials \citep{Tsakmakidis2007,Zhu2013,Cebrecos2014,Bennetts2018,wilks2022rainbow}. The interaction of waves with graded metamaterials is often termed rainbow trapping, due to the ability of the metamaterial to separate frequencies. As noted by \citet{Chaplain2020}, a better term for this interaction is rainbow reflection, since local amplifications of energy gradually decay, while trapped modes do not \citetext{also see \citealp{He2012}}.

While there are many examples of the qualitative version of the LBWA being successfully used to explain wave interaction with graded metamaterials, few studies have examined the approximation quantitatively. In particular, it is not known how the accuracy of the approximation depends on the grading strength of the metamaterial, i.e. the rate at which the resonant properties of the metamaterial vary spatially. Some discussions based on the LBWA assume that small reflections of the local Bloch wave---which can occur between adjacent unit cells with different properties---may be ignored in weakly-graded metamaterials. In other words, it is assumed that the transmission of Bloch waves is perfect except at the turning point, where instead perfect reflection occurs. \citet{Tsakmakidis2007}, who first introduced the concept of rainbow reflection in their study of an adiabatically-tapered electromagnetic waveguide, found that an excellent approximation of the wave-power transport into the device was obtained by assuming that these reflections were negligible. However, the extent to which these reflections remain negligible in other metamaterials with more significant grading is unclear.

Another key assumption of the LBWA is that decaying modes---i.e. non-propagating solutions to the Bloch-Floquet problem which are excited in graded arrays---may be ignored. \citet{Cebrecos2014}, who studied chirped sonic crystals consisting of cylinders, used coupled mode theory to approximate the array as a multilayered structure which only supports wave propagation along the direction of wave incidence. Further assuming that the dispersion relation of the crystal varies smoothly with the spatial parameter, the authors derived a system of ordinary differential equations that accurately predicts the wave amplitude envelope in the crystal. In their study of water wave interaction with line arrays of split-ring resonators with graded radii, \citet{Bennetts2018} showed that the solution was locally dominated by Rayleigh-Bloch waves, i.e. propagating homogeneous solutions to the problem of diffraction by an infinite periodic line array of scatterers \citetext{for details see \citealp{porter1999rayleigh}}. This was done by comparing the full numerical solution with an approximation, in which the velocity potential in each unit cell is represented using solely the Rayleigh-Bloch eigenfunctions of the unit cell's transfer matrix. However, none of these papers have studied the LBWA mechanistically. Here, we will investigate wave propagation in a graded metamaterial by exploring the coupling mechanism between Bloch waves in adjacent unit cells. To do this, we consider the scattering of Bloch waves across the interface between two semi-infinite uniform arrays positioned end-to-end.

The idea of using the semi-infinite problem to understand complex wave phenomena in finite arrays is not new. The present paper is inspired by the work of \citet{thompson2008new} who studied the interaction of water waves by long finite line arrays of cylinders. These authors sought to further understand the resonances of these arrays, which had previously been discussed by \citet{maniar1997wave} and are related to Rayleigh-Bloch waves. \citet{thompson2008new} decomposed the wave scattered by each cylinder into (i) the diffracted wave assuming the cylinder belongs to an infinite line array, (ii) the left and right-propagating Rayleigh-Bloch waves and (iii) decaying circular waves which originate at the array ends. The amplitudes of the Rayleigh-Bloch waves, which are excited by the incident waves at the array ends, were computed from the solution to the semi-infinite problem derived by \citet{linton2007scattering}. The so-called large array approximation of \citet{thompson2008new}, which assumes that the interactions of the decaying circular waves at the opposite array ends can be ignored, yielded numerically accurate predictions of the scattered wave coefficients. We will apply a similar modelling approach to study graded arrays of vertical barriers in a two-dimensional fluid.

In our previous work \citep{wilks2022rainbow}, we demonstrated rainbow reflection in graded arrays of surface-piercing vertical barriers of point-thickness in a two-dimensional fluid of constant and finite depth. We claimed the LBWA applied to this problem and used it to explain the rainbow reflection, although we did not verify this claim. Subsequently, we augmented the boundary value problem for the graded array of vertical barriers to include an energy absorption mechanism (i.e. damped pistons) in the inter-barrier regions. By optimising the parameters of the device, we showed that it could achieve near-perfect energy absorption over a prescribed frequency interval. Here, we will also consider the scattering of water waves by vertical barriers. This setting is advantageous for the present study because the scattering problem is simple to formulate and solve using computationally efficient methods, yet it is sufficiently non-trivial that metamaterials can support both propagating and decaying solutions to the Bloch-Floquet problem at the same frequency.

In \textsection\ref{solution_sec}, we solve the scattering of Bloch waves across the interface between two semi-infinite arrays of vertical barriers. We arrive at this solution hierarchically, after discussing (i) scattering by a single vertical barrier, (ii) scattering by finite arrays of vertical barriers, (iii) propagation of Bloch waves through infinite periodic arrays of vertical barriers and (iv) scattering by semi-infinite arrays of vertical barriers. In \textsection\ref{Bloch_coupling_sec}, we present results for the transmission of Bloch waves across the interface between two semi-infinite arrays of vertical barriers. In \textsection\ref{NILBWA}, we numerically implement the LBWA based on the assumption that the coupling of Bloch waves between adjacent subregions in a graded array behaves like the coupling across the interface between two semi-infinite arrays. Our results suggest that the LBWA accurately predicts the free surface amplitude at most frequencies, with the exception of intervals which lie just above the cutoff frequencies associated with each of the vertical barriers in the array. Specifically, these cutoff frequencies are the maximum frequency at which the corresponding infinite array problem supports propagating Bloch solutions. We argue that these errors are due to the excitation of slowly attenuating Bloch waves in the region immediately beyond the turning point.

\section{Scattering by arrays of vertical barriers}\label{solution_sec}

\subsection{A single vertical barrier}
We begin by considering the problem of wave scattering by a single surface-piercing vertical barrier, which is a classic problem in linear water wave theory \citep{Ursell1947,john1948}. Here, we consider the finite-bathymetry version of the problem. In our notation, the fluid occupies the region $\Omega=\{(x,z)|x\in\mathbb{R},-H<z<0\}\setminus \Gamma$, where $H$ is the depth, the sea-bed being situated at $z=-H$. The mean position of the free surface is situated at $z=0$ and $\Gamma=\{(0,z)|-d<z<0\}$ describes the barrier, which has submergence depth of $d$. Under the usual assumptions of time-harmonic linear water wave theory \citep{Linton2001,Mei2005}, the velocity potential of the fluid is described by $\mathrm{Re}(\phi(x,z)e^{-\mathrm{i}\omega t})$, where $t$ is time, $\omega$ is the angular frequency, and the complex valued function $\phi$ satisfies the following boundary value problem:
\begin{subequations}
\label{single_barrier_bvp}
\begin{align}
    \bigtriangleup \phi &=0 &(x,z)\in\Omega\\
    \partial_z \phi&=0 &z=-H\\
    \partial_x\phi&=0&(x,z)\in\Gamma\\
    \partial_z\phi&=\frac{\omega^2}{g}\phi&z=0\\
    \left(\frac{\partial}{\partial |x|} -\upi k_0\right)(\phi-\phi^{\mathrm{In}})&\to 0&\mbox{as\ }x\to\infty\\
    \sqrt{x^2+(z+d)^2}\|\nabla \phi\|&\to 0&\mbox{as\ }\sqrt{x^2+(z+d)^2}\to 0,
\end{align}
\end{subequations}
where $g$ is acceleration due to gravity and $\phi^{\mathrm{In}}$ is the potential of the prescribed incident wave. Moreover, $k_m$ are the solutions to the dispersion relation $k\tanh kH=\omega^2/g$, with $k_0\in\mathbb{R}^+$ and ${-\upi k_m\in((m-1)\pi/H,m\pi/H)}$ for all ${m\in\N}$. The separation of variables solution to (\ref{single_barrier_bvp}a-d) may be written as
\begin{equation}\label{sep_sol}
    \phi(x,z)=\begin{cases}
    \sum_{m=0}^\infty (A_m^{(0)}\exp(\upi k_m x)+B_m^{(0)}\exp(-\upi k_m x))\psi_m(z)&x<0\\
    \sum_{m=0}^\infty (A_m^{(1)}\exp(\upi k_m x)+B_m^{(1)}\exp(-\upi k_m x))\psi_m(z)&x<0,
    \end{cases}
\end{equation}
where the vertical eigenfunctions are
\begin{equation}
    \psi_m(z) = \left(\frac{\sinh(2k_m H)}{4k_m H}+\frac{1}{2}\right)^{-1/2}\cosh(k_m(z+H)).
\end{equation}

The expansion in \eqref{sep_sol} must also satisfy the boundary condition on the barrier and be continuously differentiable across the gap beneath the barriers. This is accomplished using the integral equation/Galerkin method of \citet{Porter1995a}, which allows us to compute accurate approximations to the coefficient vectors of the scattered field (i.e. $\mathbf{A}^{(1)}$ and $\mathbf{B}^{(0)}$) in terms of the coefficient vectors of the incident field (i.e. $\mathbf{A}^{(0)}$ and $\mathbf{B}^{(1)}$). The entries of these vectors are the coefficients in \eqref{sep_sol} after the infinite sums have been truncated at $m=N_{\mathrm{sol}}$. The method yields a scattering matrix (which we express in blockwise form) that satisfies
\begin{equation}\label{S_matrix1}
    \begin{bmatrix}S_{11}&S_{12}\\S_{21}&S_{22}\end{bmatrix}
    \begin{bmatrix}
        \mathbf{A}^{(0)}\\\ \mathbf{B}^{(1)}
    \end{bmatrix}
    =
    \begin{bmatrix}
        \mathbf{A}^{(1)}\\\ \mathbf{B}^{(0)}
    \end{bmatrix}.
\end{equation}
Other than $N_{\mathrm{sol}}$, the accuracy of the Galerkin solution depends on two other truncation parameters, which we have tuned so that at least five-figure accuracy of $B_0^{(0)}$ and $A_0^{(1)}$ is achieved across a range of test frequencies. The reader is referred to \citet{Porter1995a} for further details.

\subsection{Finite arrays of vertical barriers}\label{finite_array}
Next, we consider the scattering by $N+1$ vertical barriers which are positioned at $x=nW$ for $n\in\{0,\dots,N\}$. Although the treatment here is essentially identical to that in our earlier paper \citep{wilks2022rainbow}, we use this subsection to introduce terms which will be used subsequently in this paper. We restrict the problem to the case where the horizontal distance between barriers $W$ and their submergence depth $d$ are constant. Both restrictions could be easily relaxed (the latter by incorporating different scattering matrices at each of the different barriers) but this is not done here for ease of exposition. The solution to the multiple scattering problem may be written in the form
\begin{equation}\label{sep_sol2}
    \phi(x,z)=
    \sum_{m=0}^\infty (A_m^{(n)}\exp(\upi k_m (x-nW))+B_m^{(n)}\exp(-\upi k_m (x-nW)))\psi_m(z),
\end{equation}
for $n\in\{0,\dots,N+1\}$ and
\begin{equation}
    x\in I_n \defeq \begin{cases}
        (-\infty,0)&n=0\\
        ((n-1)W,nW)&0<n\leq N\\
        (NW,\infty)&n=N+1.
    \end{cases}
\end{equation}
A scattering matrix equation analogous to \eqref{S_matrix1}, modified to incorporate the phase-shift matrix $L=\mathrm{diag}(\exp(\upi k_m x))$, in which $0\leq m\leq N_{\mathrm{sol}}$, can be written as
\begin{equation}\label{S_matrix2}
    \begin{bmatrix}L S_{11}&L S_{12} L\\S_{21}&S_{22}L\end{bmatrix}
    \begin{bmatrix}
        \mathbf{A}^{(n-1)}\\\ \mathbf{B}^{(n)}
    \end{bmatrix}
    =
    \begin{bmatrix}
        \mathbf{A}^{(n)}\\\ \mathbf{B}^{(n-1)}
    \end{bmatrix}.
\end{equation}
for all $n\in\{1,\dots,N+1\}$. In the multiple scattering problem, $\mathbf{A}^{(0)}$ and $\mathbf{B}^{(N+1)}$ are known. To compute the remaining unknown coefficients, we use the scattering matrix method \citep{Ko1988,Bennetts2009,Montiel2015} to obtain a scattering matrix for the whole array and compute the unknown coefficients recursively. This method remains numerically stable for large $N$.

\subsection{Infinite arrays of vertical barriers}\label{infinite_barriers}
We now seek propagating solutions to the infinite array problem, i.e. the case where the barriers are positioned at $nW$ for all $n\in\mathbb{Z}$. Despite our treatment here being again essentially identical to that in our previous paper \citep{wilks2022rainbow}, we introduce terms that will be important in what follows. Our formulation and method are also analogous to those of \citet{Peter2009}. In the infinite array problem, we require that equations \eqref{sep_sol2} and \eqref{S_matrix2} hold for all $n\in\mathbb{Z}$ after redefining the horizontal intervals as $I_n=((n-1)W,nW)$. Bloch's theorem motivates seeking quasi-periodic solutions of the form
\begin{equation}\label{Bloch}
    \phi(x+nW,z)=\exp(\pm\upi q n W)\phi(x,z)
\end{equation}
for all $n\in\mathbb{Z}$, where $q$ ($-q$) is the unknown Bloch wavenumber of the forward-propagating (backward-propagating) Bloch wave. Periodicity arguments allow us to restrict $q\in(0,\pi/W)$. Solutions to \eqref{Bloch} may easily be shown to satisfy
\begin{subequations}\label{coefficient_periodicity}
    \begin{align}
    A_m^{(n)}&=\exp(\pm\upi q n W)A_m^{\pm}\\
    B_m^{(n)}&=\exp(\pm\upi q n W)B_m^{\pm},
\end{align}
\end{subequations}
i.e. the coefficients in different inter-barrier regions are related via a phase-shift which depends on the direction of the Bloch wave. Combining equations \eqref{coefficient_periodicity} and \eqref{S_matrix2} for $n=1$ yields the following generalised eigenvalue equation:
\begin{equation}\label{Bloch_matrix}
    \begin{bmatrix}L S_{11}&0_{\mathrm{M}}\\S_{21}&-I\end{bmatrix}
    \begin{bmatrix}
        \mathbf{A}^{\pm}\\\ \mathbf{B}^{\pm}
    \end{bmatrix}
    =
    \exp(\pm\upi q W)
    \begin{bmatrix}I &-L S_{12}L\\0_{\mathrm{M}}&-S_{22}L\end{bmatrix}
    \begin{bmatrix}
        \mathbf{A}^{\pm}\\ \mathbf{B}^{\pm}
    \end{bmatrix},
\end{equation}
where $I$ and $0_{\mathrm{M}}$ denotes the identity and zero matrices of dimension $N_{\mathrm{sol}}+1$, respectively. If generalised eigenvalues of \eqref{Bloch_matrix} exist (do not exist) on the unit circle at a given angular frequency $\omega$, then $\omega$ is said to lie in a passband (stopband). The eigenvectors of \eqref{Bloch_matrix} are normalised so that
\begin{equation}\label{normalisation}
    |A_0^\pm|^2-|B_0^\pm|^2+2\upi\sum_{m=1}^\infty\frac{k_m}{k_0}\mathrm{Im}(A_m^\pm \overline{B_m^\pm})=1,
\end{equation}
in which $\overline{w}$ denotes the complex conjugate of $w$. This normalisation ensures that the conservation of energy identities given in \textsection\ref{semi-infinite} and \textsection\ref{coupled_semi_infinite} take a familiar form.

\subsection{Semi-infinite arrays of vertical barriers}\label{semi-infinite}
Next, we consider the problem of scattering by a semi-infinite array of vertical barriers, which are positioned at $x=nW$ for $n\in\mathbb{N}_0$ (i.e. $n$ is a non-negative integer) and have identical submergence depth $d$. To the best of our knowledge, this particular problem has not been considered before, although the scattering of water waves by semi-infinite arrays has previously been considered in both two and three dimensions for different scatterer geometries \citep{porter2006scattering,linton2007scattering,peter2007water}. Our treatment here, which uses the so-called filtering method when Bloch waves are supported by the semi-infinite array, is based on these papers. For other solution methods to the semi-infinite diffraction problem, see \citet{martin2015one} and \citet{joseph2015reflection}.

The separation of variables expansion for the problem is given by \eqref{sep_sol2}, in which $n\in\mathbb{N}_0$ and
\begin{equation}
    I_n=\begin{cases}
        (-\infty,0)&n=0\\
        ((n-1)W,nW)&n>0.
    \end{cases}
\end{equation}
Equation \eqref{S_matrix2} then holds for all $n\in\mathbb{N}$. We first consider the case where $\omega$ is in a passband of the semi-infinite array. Anticipating the use of the solution to solve the coupling between two semi-infinite arrays, we permit the incident wave to have both left-travelling and right-travelling components. The right-travelling component is described generally by the known coefficient vector $\mathbf{A}^{(0)}$, whereas the left-travelling component is restricted to consist only of a leftward-propagating Bloch wave of known complex amplitude $\beta$. The scattered wave in $x<0$ is described by the coefficient vector $\mathbf{B}^{(0)}$. In the semi-infinite array, the scattered field consists of a rightward-propagating Bloch wave of unknown complex amplitude $\alpha$, as well as a component made up of non-propagating Bloch waves that decay as $x\to+\infty$. This motivates decomposing the coefficient vectors as
\begin{subequations}\label{filt_coeffs1}
    \begin{align}
     \mathbf{A}^{(n)} &= \alpha \exp(\upi q n W)\mathbf{A}^+ +\beta\exp(-\upi q n W)\mathbf{A}^- +\mathbf{C}^{(n)}\\
     \mathbf{B}^{(n)} &= \alpha \exp(\upi q n W)\mathbf{B}^+ +\beta\exp(-\upi q n W)\mathbf{B}^- +\mathbf{D}^{(n)}
\end{align}
\end{subequations}
for all $n\geq 1$, where $\mathbf{C}^{(n)}$ and $\mathbf{D}^{(n)}$ describe the component of the wave that decays as $x\to\infty$. This implies that $\mathbf{C}^{(n)},\mathbf{D}^{(n)}\to \mathbf{0}_{\mathrm{V}}$ as $n\to\infty$, where $\mathbf{0}_{\mathrm{V}}$ is the $(N_{\mathrm{sol}}+1)$-dimensional zero column vector.

Next, we obtain the unknown coefficients using the filtering method, which leverages information about the phase of the scattered wave (implied by the Bloch wavenumber) to greatly reduce the numerical error of the amplitude of the scattered Bloch wave compared to more direct methods \citep{linton2007scattering}. To do this, we define
\begin{subequations}\label{filt_coeffs2}
  \begin{align}
    \mathbf{\tilde{A}}^{(n)} &\defeq \begin{cases}
        \mathbf{A}^{(n)}-\beta\exp(-\upi q n W)\mathbf{A}^-&n>0\\
        \mathbf{A}^{(0)}&n=0
    \end{cases}\\
     \mathbf{\tilde{B}}^{(n)} &\defeq \begin{cases}
         \mathbf{B}^{(n)}-\beta\exp(-\upi q n W)\mathbf{B}^-&n>0\\
         \mathbf{B}^{(0)}&n=0,
     \end{cases}
\end{align}  
\end{subequations}
which approach the coefficients of the scattered Bloch wave as $n\to\infty$, i.e. $\mathbf{\tilde{A}}^{(n)}\to\alpha\exp(\upi q n W)\mathbf{A}^+$ and $\mathbf{\tilde{B}}^{(n)}\to\alpha\exp(\upi q n W)\mathbf{B}^+$. We also define
\begin{subequations}\label{filt_coeffs3}
    \begin{align}
    \mathbf{\tilde{C}}^{(n)} &\defeq \begin{cases}
        \mathbf{\tilde{A}}^{(n)}-\exp(\upi q W)\mathbf{\tilde{A}}^{(n-1)}&n>0\\
        \mathbf{A}^{(0)}&n=0
    \end{cases}\\
\mathbf{\tilde{D}}^{(n)} &\defeq \begin{cases}
        \mathbf{\tilde{B}}^{(n)}-\exp(\upi q W)\mathbf{\tilde{B}}^{(n-1)}&n>0\\
        \mathbf{B}^{(0)}&n=0,
    \end{cases}
\end{align}
\end{subequations}
which both decay to $\mathbf{0}_{\mathrm{V}}$ as $n\to\infty$. The quantities in \eqref{filt_coeffs2} and \eqref{filt_coeffs3} are related by the following telescoping sum identities:
\begin{subequations}\label{telescoping}
    \begin{align}
    \mathbf{\tilde{A}}^{(n)} &= \sum_{j=0}^n \exp(\upi q (n-j)W)\mathbf{\tilde{C}}^{(j)}\\
    \mathbf{\tilde{B}}^{(n)} &= \sum_{j=0}^n \exp(\upi q (n-j)W)\mathbf{\tilde{D}}^{(j)}.
\end{align}
\end{subequations}
Next, we substitute \eqref{telescoping} into \eqref{S_matrix2}, in order to reformulate the problem in terms of the unknowns $\mathbf{\tilde{C}}^{(n)}$ and $\mathbf{\tilde{D}}^{(n)}$. After some algebra, we eventually obtain
\begin{subequations}\label{filtering-sys}
\begin{align}
    LS_{11}\mathbf{\tilde{C}}^{(0)}+\sum_{j=0}^1 \exp(\upi q (1-j)W)[LS_{12}L\mathbf{\tilde{D}}^{(j)}-\mathbf{\tilde{C}}^{(j)}]&=\beta\exp(-\upi q W)[\mathbf{A}^- -LS_{12}L\mathbf{B}^-]\label{filtering1}\\
    S_{21}\mathbf{\tilde{C}}^{(0)}+\exp(\upi q W)[S_{22}L-I]\mathbf{\tilde{D}}^{(0)}+S_{22}L\mathbf{\tilde{D}}^{(1)}&=-\beta\exp(-\upi q W)S_{22}L\mathbf{B}^-,\label{filtering2}
\end{align}
in the $n=1$ case, and
\begin{align}
\sum_{j=0}^{n-1}\exp(\upi q(n-j)W)[(\exp(-\upi q W)LS_{11}-I)\mathbf{\tilde{C}}^{(j)}+LS_{12}L\mathbf{\tilde{D}}^{(j)}]-\mathbf{\tilde{C}}^{(n)}+LS_{12}L\mathbf{\tilde{D}}^{(n)}&=0\label{filtering3}\\
\sum_{j=0}^{n-1}\exp(\upi q(n-j)W)[\exp(-\upi q W)S_{21}\mathbf{\tilde{C}}^{(j)}+(S_{22}L-\exp(-\upi q W))\mathbf{\tilde{D}}^{(j)}]+S_{22}L\mathbf{\tilde{D}}^{(n)}&=0,\label{filtering4}
\end{align}
\end{subequations}
in the $n>1$ case. In particular, equation \eqref{Bloch_matrix} was used to show that the right-hand sides of (\ref{filtering-sys}c,d) are zero, which implies that the incident Bloch wave can only excite other wave modes at $x=0$.

To obtain the numerical solution, we assume that for some sufficiently large $P$, $\mathbf{\tilde{C}}^{(p)}=\mathbf{\tilde{D}}^{(p)}=\mathbf{0}_{\mathrm{V}}$ for all $p>P$. A $2P(N_{\mathrm{sol}}+1)$ dimensional linear system for the remaining unknown coefficients is assembled using (\ref{filtering-sys}a,b), \eqref{filtering3} for $n\in\{2,\dots,P-1\}$, \eqref{filtering4} for $n\in\{2,\dots,P\}$ and the statement $\mathbf{\tilde{C}}^{(0)}=\mathbf{A}^{(0)}$. After solving this linear system, the original desired coefficients $\mathbf{A}^{(n)}$ and $\mathbf{B}^{(n)}$ can then be obtained from \eqref{filt_coeffs1}, \eqref{filt_coeffs2} and \eqref{telescoping}. Moreover, some matrix algebra can be used to obtain the transmission and reflection operators $T_f^+$, $T_b^+$, $R_f^+$ and $R_b^+$, which satisfy
\begin{subequations}\label{semi_infinite_lin_sys}
    \begin{align}
    T_f^+\mathbf{A}^{(0)}+R_b^+\beta &= \alpha\\
    R_f^+\mathbf{A}^{(0)}+T_b^+\beta &=\mathbf{B}^{(0)},
\end{align}
\end{subequations}
in which the superscript $+$ indicates that the metamaterial occupies the positive half-line, while the subscripts $f$ and $b$ refer to the forward and backward scattering problems, respectively. In \eqref{semi_infinite_lin_sys}, $T_f^+$ is a $(N_{\mathrm{sol}}+1)$-dimensional row vector, $R_b^+$ is a scalar, $R_f^+$ is a $(N_{\mathrm{sol}}+1)$-dimensional square matrix and $T_b^+$ is a $(N_{\mathrm{sol}}+1)$-dimensional column vector. If only propagating modes are incident to the system, \eqref{semi_infinite_lin_sys} can be reduced to a $2\times2$ matrix equation of the form
\begin{equation}\label{semi_infinite_lin_sys_reduced}
    S(0,d)\begin{bmatrix}
        A_0^{(0)}\\ \beta
    \end{bmatrix}=
    \begin{bmatrix}
        \alpha \\ B_0^{(0)}
    \end{bmatrix},
\end{equation}
i.e. it relates the amplitudes of the propagating waves. The reason for using the notation $S(0,d)$ for the $2\times2$ matrix will become apparent in \textsection\ref{coupled_semi_infinite}.

Next, we consider conservation of energy. In the forward scattering problem, we set $A_m^{(0)}=\delta_{0m}$ (in which $\delta_{ij}$ denotes the Kronecker delta) and $\beta=0$. The reflection and transmission coefficients are then given by $R=B_0^{(0)}$ and $T=\alpha$. In the backward scattering problem, we set $\beta=1$ and $A_m^{(0)}=0$ for all $0<m<N_{\mathrm{sol}}$. In this case, the reflection and transmission coefficients are given by $T=B_0^{(0)}$ and $R=\alpha$. In both scattering problems, the conservation of energy identity $|R|^2+|T|^2=1$ can be shown to hold by applying Green's second identity. The familiar form of this identity is guaranteed by the normalisation of the Bloch waves given in \eqref{normalisation}.

We remark that when $\omega$ is in a stopband, no propagating Bloch waves exist in the metamaterial region and the filtering method is not required. In this case, the only required diffraction matrix is $R_f^+$, which can be approximated accurately using the solution for a finite, regularly spaced array of vertical barriers, that is, the method outlined in \textsection\ref{finite_array}. In the absence of propagating Bloch waves, all of the matrices $T_f^+$, $T_b^+$ and $R_b^+$ are not well defined. For consistency of notation in this case, we write a $1\times 1$ scattering matrix equation of the form $S(0,d)A_0^{(0)}=B_0^{(0)}$. The conservation of energy identity for the forward scattering problem in which $A_m^{(0)}=\delta_{0m}$ is simply $|R|=1$, where $R= B_0^{(0)}$.

We also require the solution to the problem where the barriers occupy the negative half-line. Specifically, the barriers are situated at $x=-nW$ for $n\in\mathbb{N}$. This solution is obtained by applying the transformation $x\mapsto-(x+W)$. Following some lengthy algebra, we obtain the following relationships between the diffraction matrices for the positive and negative half-line problems:
\begin{subequations}
    \begin{align}
    R_b^- &= LR_f^+L&T_b^-&=JT_f^+L\\
    T_f^+&=LT_b^-J&R_f^- &=J^2R_b^+,
\end{align}
\end{subequations}
where the number $J$ has been defined as
\begin{equation}
    J\defeq\exp(-\upi k_m W)\frac{A_m^+}{B_m^-}=\exp(\upi k_m W)\frac{B_m^+}{A_m^-},
\end{equation}
which can be shown to be independent of the choice of $m$. The negative half-line diffraction matrices satisfy
\begin{align}\label{negative_halfline_array}
    T_f^-\gamma + R_b^-\mathbf{B}^{(0)}&=\mathbf{A}^{(0)}\\
    R_f^-\gamma + T_b^-\mathbf{B}^{(0)}&=\eta
\end{align}
where $\gamma$ and $\eta$ denote the amplitudes of the incident (right-travelling) and scattered (left-travelling) Bloch waves in the metamaterial region, respectively. Similarly to the right semi-infinite array problem, equation \eqref{negative_halfline_array} may be reduced to a $2\times 2$ scattering matrix equation of the form
\begin{equation}
    S(d,0)\begin{bmatrix}
        \gamma\\ B_0^{(0)}
    \end{bmatrix}
    =
    \begin{bmatrix}
        A_0^{(0)}\\ \eta
    \end{bmatrix},
\end{equation}
when evanescent modes are discarded.

\subsection{Two coupled semi-infinite arrays}\label{coupled_semi_infinite}
\begin{figure}
    \centering
    \includegraphics[width=0.8\textwidth]{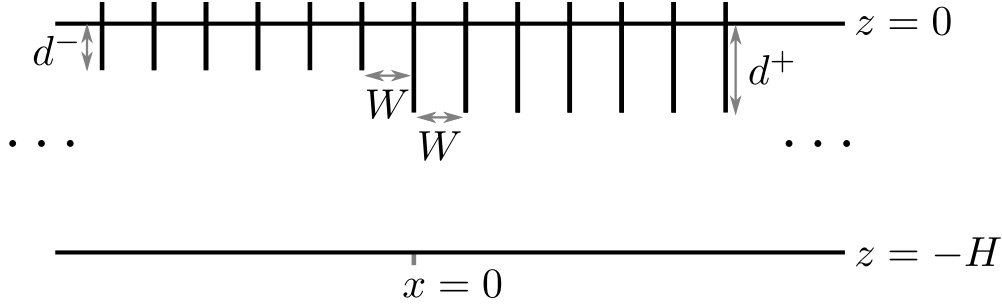}
    \caption{A schematic of the geometry of the problem of two coupled semi-infinite arrays of surface-piercing vertical barriers in a fluid of finite depth.}
    \label{fig:schematic}
\end{figure}

Lastly, we solve the problem of two coupled semi-infinite arrays. The submergence depth of the barriers of the semi-infinite array on the negative (positive) half-line is denoted $d^-$ ($d^+$). For simplicity we will only consider the case when the spacing $W$ is identical for both semi-infinite arrays, although this condition would be easy to relax at the cost of introducing more notation. A schematic of the geometry is given in figure \ref{fig:schematic}.

In general, we consider the case where incident Bloch waves can originate from both semi-infinite arrays. The Bloch wave incident from the left (right) semi-infinite array has complex amplitude $\alpha^-$ ($\beta^+$) and propagates towards the right (left). The unknown amplitude of the scattered wave travelling left (right) in the left (right) semi-infinite array is denoted $\beta^-$ ($\alpha^+$). With reference to the diffraction matrices in \textsection\ref{semi-infinite}, these unknown amplitudes can be obtained from the following linear system:
\begin{equation}\label{2_semi_infinite_system1}
    \begin{bmatrix}
        -1&0&T_f^+&\mathbf{0}_{\mathrm{V}}^\intercal\\
        \mathbf{0}_{\mathrm{V}}&\mathbf{0}_{\mathrm{V}}&R_f^+&-I\\
        \mathbf{0}_{\mathrm{V}}&\mathbf{0}_{\mathrm{V}}&-I&R_b^-\\
        0&-1&\mathbf{0}_{\mathrm{V}}^\intercal&T_b^-
    \end{bmatrix}
    \begin{bmatrix}
        \alpha^+\\ \beta^-\\ \mathbf{A}^{(0)}\\ \mathbf{B}^{(0)}
    \end{bmatrix}
    =\begin{bmatrix}
        -R_b^+\beta^+\\-T_b^+\beta^+\\-T_f^- \alpha^-\\ -R_f^- \alpha^-
    \end{bmatrix},
\end{equation}
in which $\intercal$ denotes the transpose. When $\alpha^-=1$ and $\beta^+=0$, the reflection and transmission coefficients for this problem are $R=\beta^-$ and $T=\alpha^+$. Green's second identity and the Bloch wave normalisation \eqref{normalisation} yields the conservation of energy identity $|R|^2+|T|^2=1$. When $\alpha^-=0$ and $\beta^+=1$, the aforementioned identity again holds with the roles of $R$ and $T$ reversed. After inverting the matrix in \eqref{2_semi_infinite_system1}  and using matrix algebra to eliminate $\mathbf{A}^{(0)}$ and $\mathbf{B}^{(0)}$, the following $2\times2$ linear system for the unknown Bloch amplitudes is obtained:
\begin{equation}\label{2_semi_infinite_system1_reduced}
    S(d^-,d^+)\begin{bmatrix}
        \alpha^-\\ \beta^+
    \end{bmatrix}=\begin{bmatrix}
        \alpha^+\\ \beta^-
    \end{bmatrix}.
\end{equation}

The previous discussion assumed that propagating Bloch waves are defined in both semi-infinite regions. If $\omega$ is in the stopband of the right semi-infinite array, then \eqref{2_semi_infinite_system1} reduces to
\begin{equation}\label{2_semi_infinite_system2}
    \begin{bmatrix}
       \mathbf{0}_{\mathrm{V}}&R_f^+&-I\\
       \mathbf{0}_{\mathrm{V}}&-I&R_b^-\\
       -1&\mathbf{0}_{\mathrm{V}}^\intercal&T_b^-
    \end{bmatrix}
    \begin{bmatrix}
        \beta^-\\ \mathbf{A}^{(0)}\\ \mathbf{B}^{(0)}
    \end{bmatrix}
    =\begin{bmatrix}
        \mathbf{0}_{\mathrm{V}}\\-T_f^- \alpha^-\\ -R_f^- \alpha^-
    \end{bmatrix}.
\end{equation}
For $\alpha^-=1$, the conservation of energy condition becomes $|R|=1$, where $R=\beta^-$. For notational consistency, we remark that \eqref{2_semi_infinite_system2} reduces to the following $1\times 1$ linear system:
\begin{equation}\label{2_semi_infinite_system2_reduced}
    S(d^-,d^+)\begin{bmatrix} \alpha^-
    \end{bmatrix}=\begin{bmatrix}
        \beta^-
    \end{bmatrix}.
\end{equation}

\section{Bloch wave coupling in semi-infinite arrays}\label{Bloch_coupling_sec}
In the remainder of this paper, we fix the parameters $H=20$\,m and $W=2$\,m. With regards to the numerical parameters, we set the truncation of the solution expansion $N_{\mathrm{sol}}=20$. Further, the point in the array where decaying effects are assumed to have vanished is fixed at $P=50$ for the remainder of this paper. This value of $P$ was chosen so that least four figures of convergence was obtained for $\alpha$, where $\alpha$ is the amplitude of the right-travelling Bloch wave in the problem of forward scattering by a single semi-infinite array described in \textsection\ref{semi-infinite}. The validation of our method for the problem of two coupled semi-infinite arrays is discussed in the next subsection.

\subsection{Method validation}\label{validation_sec}
Several steps were taken to validate the method given in \textsection\ref{coupled_semi_infinite} and its subsequent implementation. First, the absolute error of the conservation of energy identities was checked and found to be less than $5 \times 10^{-7}$ across the range of parameters and frequencies considered here. Second, we chose an arbitrary selection of barriers on which to verify the boundary and matching conditions. We confirmed that to a good approximation (i) the normal derivative of the potential on each barrier was zero and (ii) the potential was continuously differentiable beneath the barrier. Third, we recognise that the solution to the two coupled semi-infinite arrays problem must reduce to the infinite array problem when the barriers of both arrays have the same submergence depth, i.e. $d^+=d^-$. This means that Bloch waves should not undergo any interaction at the interface and that $S(d^-,d^+)$ is the $2\times 2$ identity matrix. This has been verified.

\subsection{Transmission between two semi-infinite arrays}
We first recall some results from \citep{wilks2022rainbow} about Bloch wave dispersion in an infinite array of vertical barriers. Periodic arrays of vertical barriers have a single finite passband at low frequencies, which lies beneath an infinite stopband. The so-called \textit{cutoff frequency} which separates the passband and stopband is closely related to the resonant frequency of a pair of vertical barriers, which decreases as the submergence depth of the two barriers increases \citep{wilks2022rainbow}.

\begin{figure}
    \centering
    \includegraphics[width=0.8\textwidth]{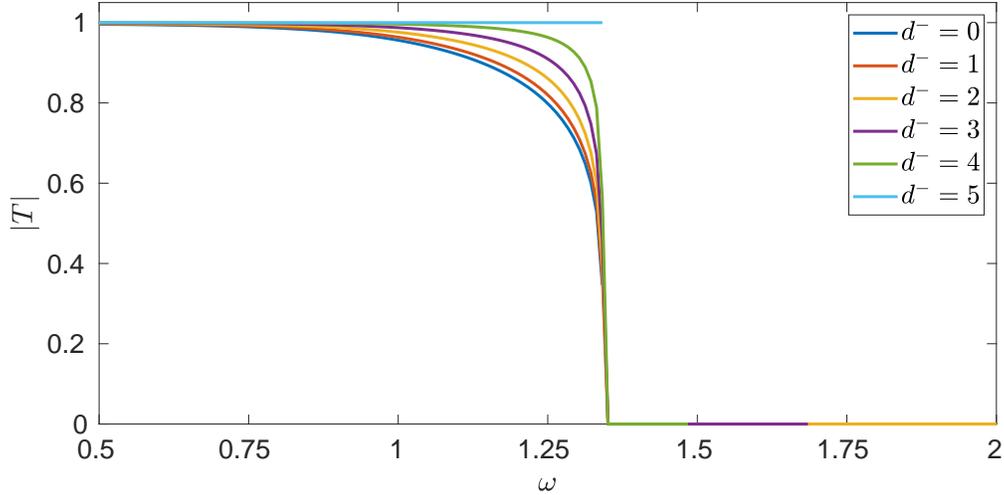}
    \caption{The effect of frequency on the transmission of Bloch waves from the left semi-infinite array to the right semi-infinite array. The submergence depth of the barriers in the right semi-infinite array is fixed at $d^+=5$\,m, while the submergence depth of the barriers in the left semi-infinite array ranges from $d^-=1$\,m to $d^-=5$\,m. The perfect transmission indicated by the curve for $d^+=d^-$ provides validation for the method of coupling two semi-infinite arrays---see the discussion in \textsection\ref{validation_sec} for further details. When $d^-=0$\,m, the problem reduces to scattering by a single semi-infinite array---the corresponding transmission curve is also included.}
    \label{fig:bloch_trans_variable_depth}
\end{figure}

Curves showing the effect of frequency on the transmission of Bloch waves from the left semi-infinite array to the right semi-infinite array are given in figure \ref{fig:bloch_trans_variable_depth}. In that figure, we fix the submergence depth of the barriers in the right semi-infinite array to be $d^+=5$\,m, while varying the submergence depth of the barriers in the left semi-infinite array from $d^-=0$\,m to $d^-=5$\,m. We observe that at low frequencies, transmission is very close to unity regardless of the submergence depth of the barriers in the left semi-infinite array. This is because the interaction between the barriers and the wave is negligible when $k_0d\ll 1$ and $d\ll H$ \citep{Ursell1947}. For $d^-<d^+$, transmission decreases as the frequency approaches the cutoff frequency from below. The rate of this decrease is higher when $d^+-d^-$ is larger. When $0<d^-<5$ and $\omega$ is in the stopband of the right semi-infinite array (approximately $1.342$\,s$^{-1}$), Bloch waves cannot propagate to the right, thus $T$ vanishes as only reflection can occur. In these cases, $\omega$ eventually enters the stopband of the left semi-infinite array as well. When this happens, the problem is not well defined and the transmission curve terminates, although several of these terminations occur outside of the frame of figure \ref{fig:bloch_trans_variable_depth}. When $d^-=0$\,m, the open sea region $x<0$ supports propagating plane waves at all frequencies, thus the corresponding transmission curve never terminates.

The results from this section have implications for the LBWA. In any unit cell of a graded array, the local Bloch waves must couple with the local Bloch waves in adjacent unit cells. It is reasonable to expect that this coupling will behave like the coupling of Bloch waves across two semi-infinite arrays. When the grading is not sufficiently weak, the difference between the submergence depths of adjacent barriers may not be small. This means that the naive version of the LBWA, which treats Bloch wave coupling before the turning point as perfect transmission, may not apply. Instead, these reflections should be accounted for and the problem should be treated like a multiple scattering problem. In the next section, we describe and implement a version of the LBWA which accounts for reflections of Bloch waves yet continues to ignore the contributions of decaying Bloch modes.

\section{Numerical implementation of the LBWA}\label{NILBWA}
We consider the problem of scattering by a finite graded array of $N$ vertical barriers using an approximate method based on the LBWA. In particular, we assume that between every adjacent pair of barriers, the wave can be represented as the superposition of right and left travelling Bloch waves. This assumption reduces the problem to the well known problem of one-dimensional multiple scattering of waves on a string  \citetext{see \citealp{martin2014n} for an outline of these problems}. In this analogy, the grading of the submergence depths of the barriers in the water wave metamaterial would be equivalent to the string having piecewise-constant material properties. To describe our scattering problem in this way, we must obtain the transmission and reflection coefficients that couple Bloch waves in adjacent regions---we derive these from the coupling of two semi-infinite arrays as discussed in \textsection\ref{coupled_semi_infinite}. The solution obtained using this method will be contrasted with the solution obtained using the semi-analytic method introduced in \textsection\ref{finite_array}.

\subsection{Method}\label{lbwa_method}
As in \textsection\ref{finite_array}, the barriers are positioned at $x=nW$ and have submergence depth $d^{(n)}$ for $n\in\{0,\dots,N\}$. In the region $x<0$, we assume a plane incident wave of the form $\exp(\upi k_0 x)\psi_0(z)$ and also approximate the scattered field as a plane wave, which implies that
\begin{subequations}\label{local_Bloch_coeffs1}
    \begin{align}
    A_m^{(0)}=\tilde{A}_\mathrm{Inc}\delta_{0m}\\
    B_m^{(0)}\approx\tilde{B}_\mathrm{R}\delta_{0m},
\end{align}
where $\tilde{B}_\mathrm{R}$ is unknown and
\begin{equation}\nonumber
    \tilde{A}_{\mathrm{Inc}}\approx A_{\mathrm{Inc}}\frac{g}{\omega}\left(\frac{\sinh(2k_0H)}{4k_0H}+\frac{1}{2}\right)^{1/2},
\end{equation}
where $A_{\mathrm{Inc}}$ is the amplitude of the incident wave. To the right of the array, the transmitted wave is approximated as a plane wave of the form
\begin{equation}
    A_m^{(N+1)} \approx \tilde{A}_\mathrm{T}\delta_{0m},
\end{equation}
\end{subequations}
where $\tilde{A}_\mathrm{T}$ is unknown. For $x\in(nW,(n+1)W)$ and $0\leq n <N$, the wave is approximated as the superposition of Bloch waves that correspond to the infinite array where the barriers have submergence $d^{(n)}$ \textit{if solutions to the Bloch-Floquet problem exist}. That is, we assume that the potential is a superposition of right and left travelling Bloch waves with complex amplitudes $\alpha^{(n)}$ and $\beta^{(n)}$, respectively. This implies
\begin{subequations}\label{local_Bloch_coeffs2}
    \begin{align}
    A_m^{(n)}&\approx\alpha^{(n)}\exp(\upi q^{(n-1)} n W)\mathbf{A}^{(n-1)+}+\beta^{(n)}\exp(-\upi q^{(n-1)} n W)\mathbf{A}^{(n-1)-}\\
    B_m^{(n)}&\approx\alpha^{(n)}\exp(\upi q^{(n-1)} n W)\mathbf{B}^{(n-1)+}+\beta^{(n)}\exp(-\upi q^{(n-1)} n W)\mathbf{B}^{(n-1)-},
\end{align}
\end{subequations}
where $q^{(j)}$ is the Bloch wavenumber and $\mathbf{A}^{(j)\pm}$ and $\mathbf{B}^{(j)\pm}$ are the Bloch wave coefficients for the infinite array of barriers with submergence depth $d^{(j)}$. These quantities are computed using the method given in \textsection\ref{infinite_barriers}. If the Bloch-Floquet problem has no solutions for the infinite array where the barriers have submergence $d^{(n)}$, then we assume $A_m^{(n)}=B_m^{(n)}=0$.

To compute the LBWA, we must consider three cases:
\subsubsection{Case 1: total reflection by the first barrier}
This case governs the situation where the region $0<x<W$ does not support propagating Bloch waves, that is, the frequency is in the stopband of the infinite array where the barriers have submergence $d^{(0)}$. The LBWA suggests that there is no wave amplitude anywhere in the array and that transmission by the array is zero. The interaction at the first barrier is modelled using the single semi-infinite array diffraction problem where the barriers occupy the positive half-line. Thus we need only compute $\tilde{B}_\mathrm{R}=S(0,d^{(0)})\tilde{A}_\mathrm{Inc}$.

\subsubsection{Case 2: total reflection within the graded array}
This case describes the situation where the local Bloch wave is totally reflected inside the array. Let $p$ be the smallest value in $\{1,\dots,N\}$ such that $\omega$ is in the stopband of the infinite array of barriers with submergence depth $d^{(p)}$. This implies that $x=pW$ is the turning point and no Bloch waves exist past this point. At the leftmost barrier of the array, we assume that the wave interaction can be locally modelled using the single semi-infinite array diffraction problem. Thus, we use \eqref{semi_infinite_lin_sys_reduced} to relate the amplitudes of plane waves to the left and Bloch waves to the right of the leftmost barrier, i.e.
\begin{subequations}\label{local_Bloch_system}
\begin{equation}
    S(0,d^{(0)})\begin{bmatrix}
        \tilde{A}_{\mathrm{Inc}}\\ \beta^{(1)}
    \end{bmatrix}=\begin{bmatrix}
        \alpha^{(1)}\\ \tilde{B}_R
    \end{bmatrix}.
\end{equation} For $2\leq n\leq p$, we assume that the wave interaction can be locally modelled using the two coupled semi-infinite arrays problem. This problem must be transformed so that the first barrier of the right semi-infinite array occurs at $x=(n-1)W$ instead of at $x=0$. By combining \eqref{2_semi_infinite_system1_reduced} with the phase shift matrices $P_n=\mathrm{diag}[\exp(-\upi q^{(n-1)}nW),\exp(\upi q^{(n-2)}nW)]$ and $Q_n=\mathrm{diag}[\exp(-\upi q^{(n-1)}nW),\exp(\upi q^{(n-2)}nW)]$, we obtain
\begin{equation}
    P_nS(d^{(n-2)},d^{(n-1)})Q_n\begin{bmatrix}
        \alpha^{(n-1)}\\ \beta^{(n)}
    \end{bmatrix}=\begin{bmatrix}
        \alpha^{(n)}\\ \beta^{(n-1)}
    \end{bmatrix}.
\end{equation}
Lastly, the amplitudes of the Bloch waves in the region immediately preceding the turning-point barrier are again related by assuming that the wave interaction can be locally modelled using the two coupled semi-infinite arrays problem. However, in this problem no Bloch wave is defined in the right semi-infinite array. Thus we use \eqref{2_semi_infinite_system2_reduced} to model the reflection of Bloch waves in the left semi-infinite array, which after transformation becomes
\begin{equation}\label{turning_point_reflection}
    \exp(2\upi q^{(p-1)}pW)S(d^{(p-1)},d^{(p)})\alpha^{(p)}=\beta^{(p)}.
\end{equation}
\end{subequations}
Equations (\ref{local_Bloch_system}a-c) form a system of $2p+1$ equations and $2p+1$ unknowns, which can be solved using standard methods. The approximation of the fluid potential can be recovered using (\ref{local_Bloch_coeffs1}a,b) and (\ref{local_Bloch_coeffs2}a,b).

\subsubsection{Case 3: partial transmission}\label{case3}
This last case governs the case where local Bloch waves are defined throughout the array and therefore the array can transmit energy. The solution proceeds similarly to case 2, where (\ref{local_Bloch_system}a) holds and (\ref{local_Bloch_system}b) holds for $2\leq n \leq N$. However, (\ref{local_Bloch_system}c) no longer applies. Instead, the interaction of the Bloch waves to the left of the $(N+1)$st barrier with the plane wave to the right is modelled using the semi-infinite array diffraction problem where the barriers occupy the negative half-line. This implies that
\begin{equation}
    P_{N+1}S(d^{(N)},0)Q_{N+1}\begin{bmatrix}
        \alpha^{(N)}\\ 0
    \end{bmatrix}
    =
    \begin{bmatrix}
        \tilde{A}_\mathrm{T} \\ \beta^{(N)}
    \end{bmatrix}
\end{equation}
where the phase-shift matrices have been defined as
\begin{equation*}
    P_{N+1}=\mathrm{diag}[1,\exp(\upi q^{(N)}(N+1)W]\quad\mbox{and}\quad Q_{N+1}=\mathrm{diag}[\exp(\upi q^{(N)}(N+1)W,1].
\end{equation*}
Case 3 requires that $d^{(N-1)}=d^{(N)}$ so that Bloch waves can be consistently defined in the region $x\in((N-1)W,NW)$.

\subsection{Validation}
\begin{figure}
    \centering
    \includegraphics[width=\textwidth]{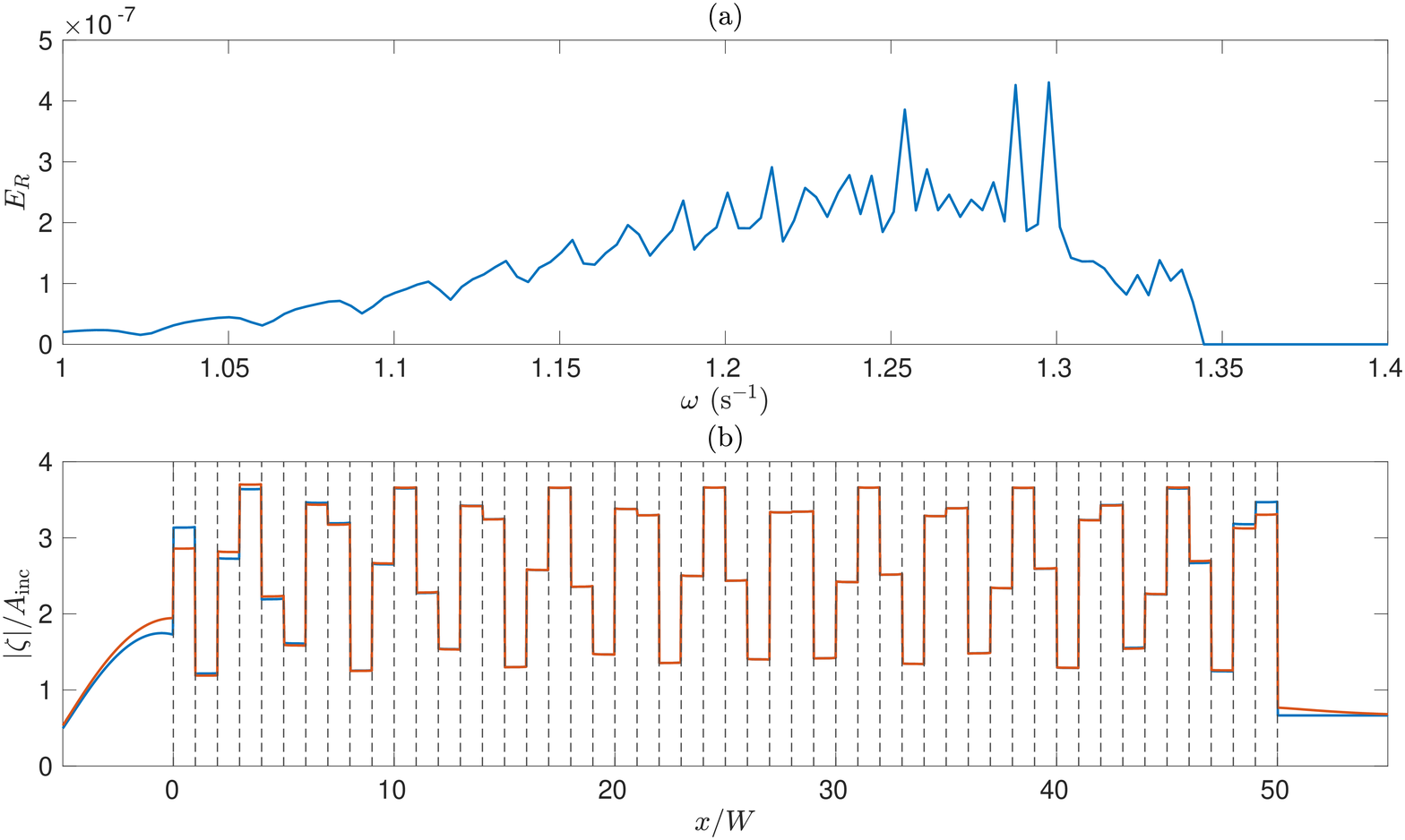}
    \caption{Validation of the LBWA for a finite array of vertical barriers for which $N=50$ and $d^{(n)}=5$\,m for all $0\leq n\leq N$. Panel (a) displays the absolute error of the reflection coefficient. Panel (b) contrasts the absolute value of the free surface elevation computed using the LBWA (blue line) and using the semi-analytic method (red line) at the angular frequency $\omega=1.2$\,s$^{-1}$. The $x$-coordinates of the vertical barriers are marked with dashed black lines.}
    \label{fig:long_array_approx}
\end{figure}

We will not validate our implementation of the LBWA using a direct comparison with the semi-analytic solution for graded arrays, since the origin of any discrepancy would be unclear. In particular, these could be fundamental to the LBWA or they could originate from mathematical or coding mistakes. To rule out the latter, we consider a long array of regularly-spaced and identical vertical barriers, since the LBWA should perform well in this case. Indeed, the LBWA reduces to a long array approximation analogous to that of \citet{thompson2008new}, as it assumes that only propagating Bloch waves excited at each of the array ends can interact at the opposite ends. In other words, the evanescent modes excited at each end do not interact. To evaluate the LBWA over a range of frequencies, we consider the absolute error of the reflection coefficient defined as $E_R = |\tilde{R}-R|$, where $\tilde{R}$ is computed from the LBWA (i.e. it is the value of $\tilde{B}$ when $\tilde{A}_{\mathrm{Inc}}=1$) and $R$ is computed using the semi-analytic method described in \textsection\ref{finite_array}. In figure \ref{fig:long_array_approx}, results are presented for a finite array of $N=50$ barriers with submergence depth $d^{(n)}=5$\,m for all $0\leq n\leq N$. In panel (a), we observe that the absolute error of the reflection coefficient remains bounded below $5\times 10^{-7}$ below the cutoff frequency $\omega\approx 1.342$\,s$^{-1}$ and is numerically zero above the cutoff. 

We also assess how well the LBWA predicts local energy amplification within the array. To do this, we use the complex-valued free surface elevation given by 
\begin{equation}
    \zeta(x) = \frac{\upi\omega}{g}\phi(x,0).
\end{equation}
In figure \ref{fig:long_array_approx}(b), the absolute value of the free surface elevation at $\omega=1.2$\,s$^{-1}$ is computed using both the LBWA (blue line) and the semi-analytic solution (red line) for the array. We observe that there is excellent agreement between the two solutions across most of the horizontal domain. Small errors occur in the neighborhood of the zeroth and $N$th barriers, which are due to evanescent modes excited at the array ends that are not considered by the LBWA. These results are completely analogous to those of \citet{thompson2008new}. We conclude that our method performs as expected for a problem that the LBWA should describe accurately. This suggests that any errors produced by our method in other problems are fundamental to the LBWA.

\subsection{Results}\label{NILBWA_results}
\begin{figure}
    \centering
    \includegraphics[width=\textwidth]{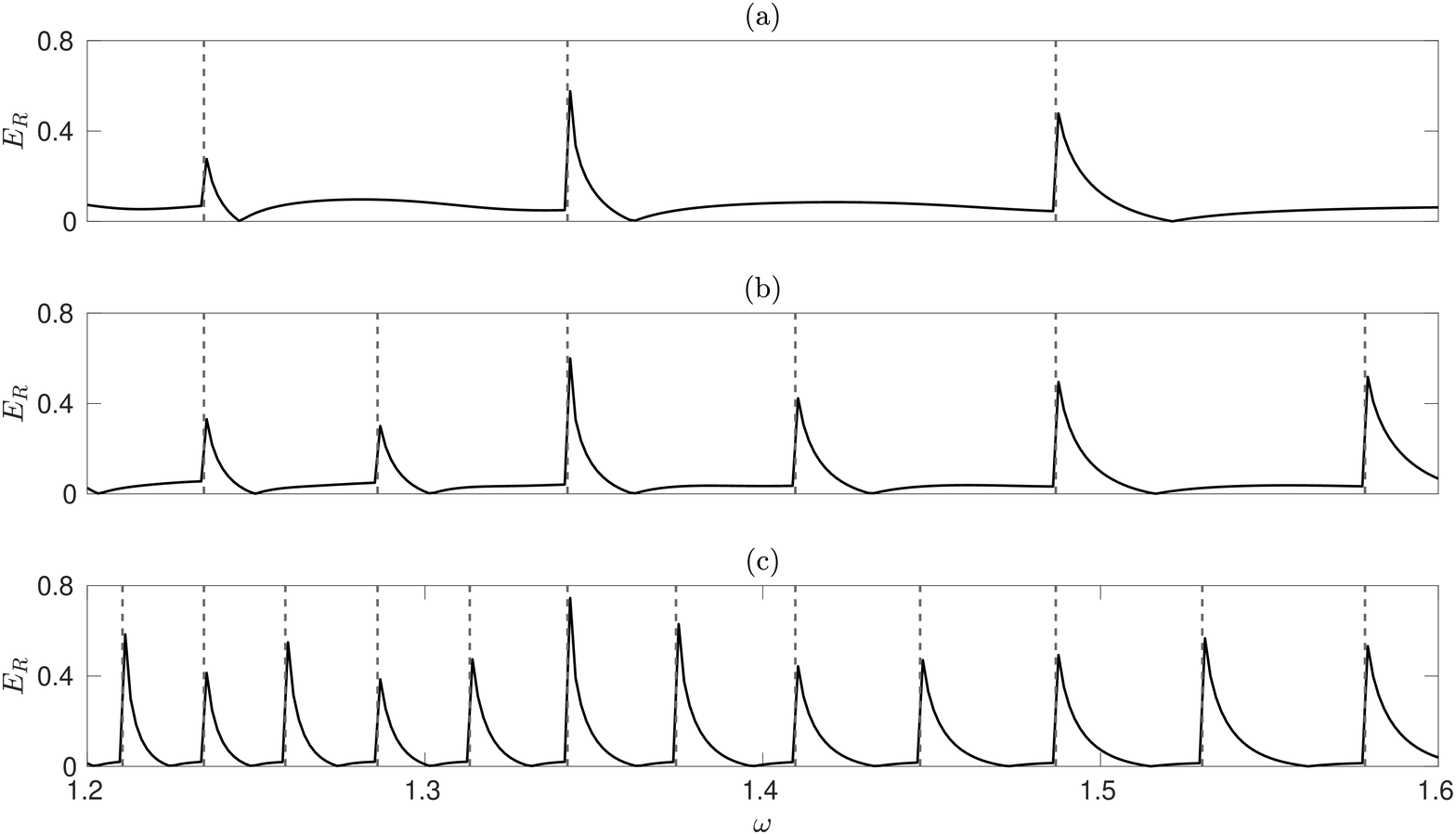}
    \caption{The absolute error of the reflection coefficient for arrays of vertical barriers parameterised by (a) $\Delta d=1$\,m, (b) $\Delta d=0.5$\,m and (c) $\Delta d=0.25$\,m. The cutoff frequencies $\omega^{(n)}$ are indicated with dashed vertical lines. Specifically, these are the cutoff frequencies for an infinite array of vertical barriers with spacing $W$ and submergence $d^{(n)}$, for $0\leq n \leq N$.}
    \label{fig:local_bloch_error}
\end{figure}

We consider linearly-graded arrays of $N$ vertical barriers described by $\Delta d = d^{(n)}-d^{(n-1)}$, where we restrict $\Delta d$ so that $N+1=d^{(N)}/\Delta d\in\mathbb{N}$. Moreover, we fix $d^{(N)}=10$\,m which implies that $d^{(0)}=\Delta d$. Three such arrays are considered, for which the grading parameters are $\Delta d=0.25$\,m, $\Delta d=0.5$\,m and $\Delta d = 1$\,m. In this subsection, we restrict our investigation to case 2, i.e. we only consider frequencies at which the LBWA predicts total reflection within the array. Plots showing the absolute error of the reflection coefficient for each of these arrays are given in figure \ref{fig:local_bloch_error}. We observe that the curves of $E_R$ are sawtooth-like. The peaks of these error curves occur just above the angular frequencies $\omega^{(n)}$, where $\omega^{(n)}$ is the cutoff frequency of the infinite array of vertical barriers with submergence depth $d^{(n)}$. Away from these peaks, the error approaches a baseline that is lower for more weakly-graded arrays. The error peaks for more weakly-graded arrays are more densely packed because these arrays contain more different barrier submergence depths, i.e they have more internal cutoff frequencies. 

\begin{figure}
    \centering
    \includegraphics[width=\textwidth]{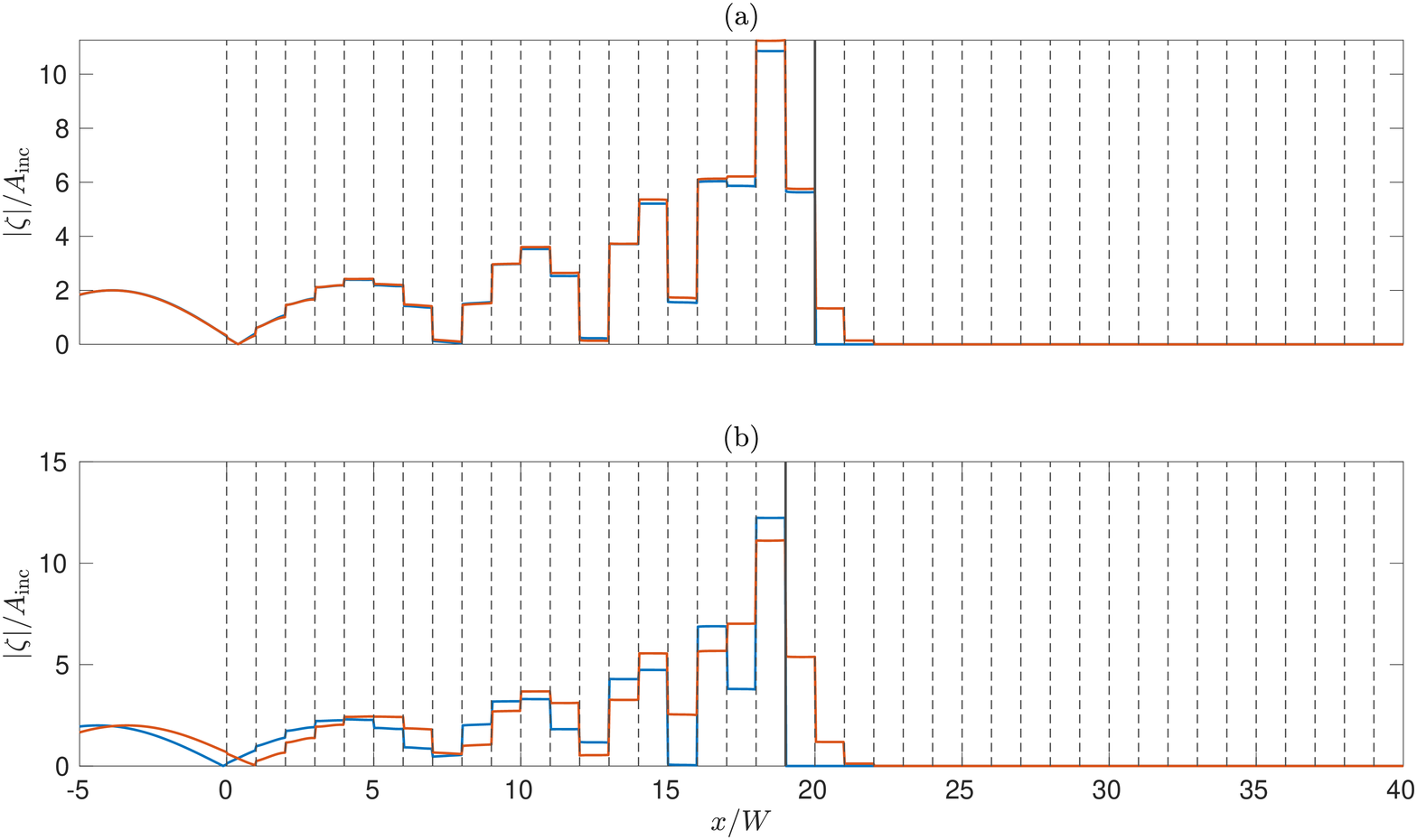}
    \caption{The free surface elevation in the graded array of vertical barriers with grading parameter $\Delta d=0.25$\,m and angular frequencies (a) $\omega=1.341$\,s$^{-1}$ and (b) $\omega=1.343$\,s$^{-1}$. A complete description of the array geometry is provided in \textsection\ref{NILBWA_results}. The elevation curves were computed using the semi-analytic method (red line) and our implementation of the LBWA (blue line). The positions of the vertical barriers (i.e.  $x=nW$ for $0\leq n \leq N$) are marked with dashed black lines and the position of the turning point (i.e. $x=pW$) is marked with a solid black line.}
    \label{fig:approximation_comparison_weak}
\end{figure}

\begin{figure}
    \centering
    \includegraphics[width=\textwidth]{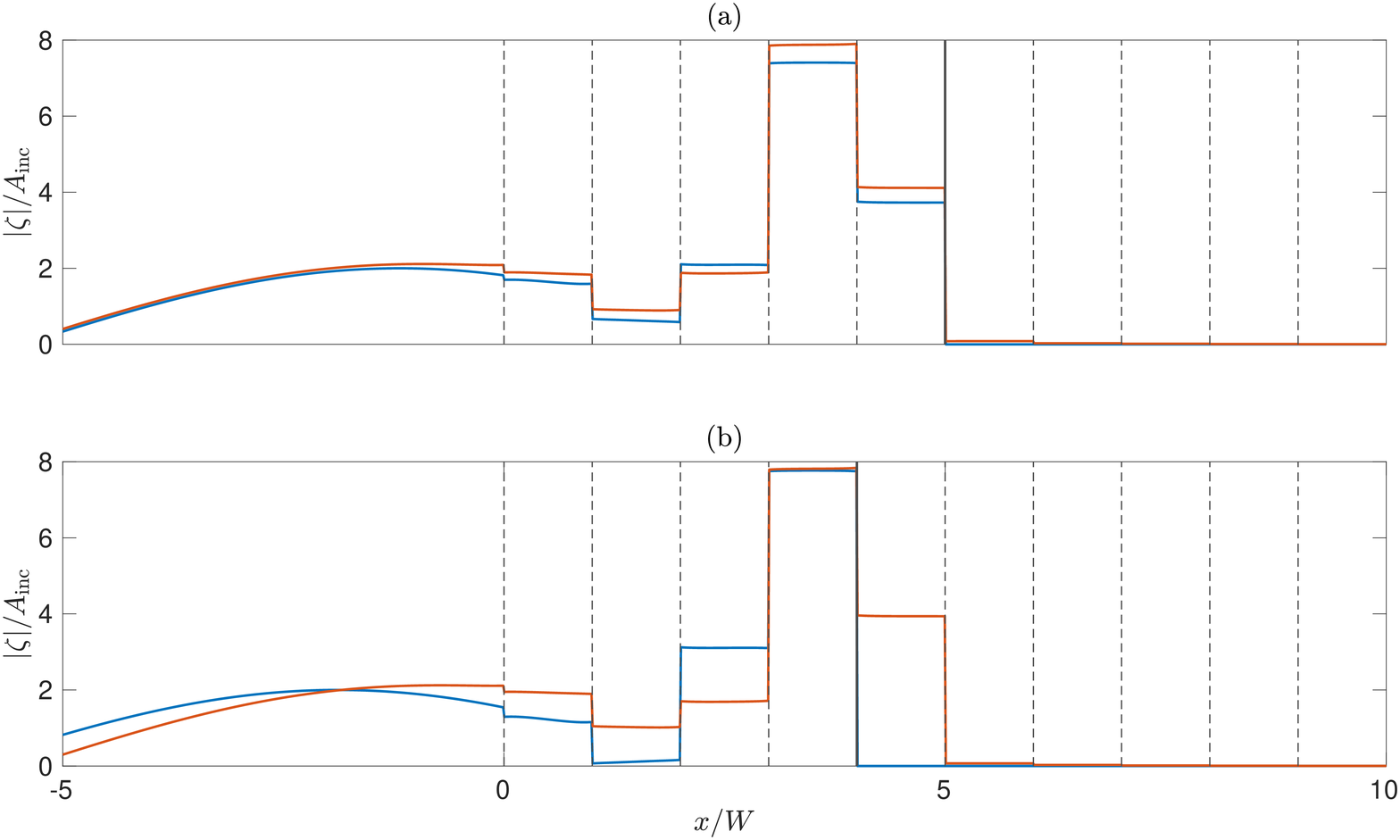}
    \caption{As for figure \ref{fig:approximation_comparison_weak} but with $\Delta d = 1$\,m.}
    \label{fig:approximation_comparison_strong}
\end{figure}

Next, we seek to explain why the cutoff frequencies induce peaks in the error curves. To do this, we use the absolute value of the free surface to compare the LBWA and semi-analytic solutions for the graded array with $\Delta d=0.25$\,m (figure \ref{fig:approximation_comparison_weak}) and $\Delta d=1$\,m (figure \ref{fig:approximation_comparison_strong}). We compare the solutions at $\omega=1.341$\,s$^{-1}$ (panel (a) in each figure) and $\omega=1.343$\,s$^{-1}$ (panel (b) in each figure). The former angular frequency is just below the cutoff frequency $\omega^{(j)}$ associated with the barrier with submergence depth $d^{(j)}=5$\,m, whereas the latter angular frequency is just above this cutoff. The index $j$ of this barrier depends on $\Delta d$. Specifically, $j=19$ in figure \ref{fig:approximation_comparison_weak} and $j=4$ in figure \ref{fig:approximation_comparison_strong}.

In figures \ref{fig:approximation_comparison_weak}(a) and \ref{fig:approximation_comparison_strong}(a), which correspond to the case where $\omega$ is just below $\omega^{(j)}$, we observe excellent agreement between the solutions for $x<pW$, where $p=j+1$ is the turning point. In results not shown here, similarly-good agreement was found across a range of different array configurations and frequencies, provided that the frequency is away from the peaks in $E_R$. This suggests that decaying Bloch waves are negligible in these cases. 

We briefly discuss the small discrepancy in figure \ref{fig:approximation_comparison_weak}(a) in the region $pW<x<(p+1)W$. As this region does not support propagating Bloch modes, our implementation of the LBWA defines the amplitude in this region to be zero. However, this region does support decaying Bloch waves. To explain why this error does not significantly affect the accuracy of the LBWA for $x<pW$, we note that the reflection coefficient used by the LBWA at $x=pW$ was computed from the version of the two coupled semi-infinite arrays problem in which there is no propagating mode in the right array. This means that the decaying Bloch wave in $pW<x<(p+1)W$ is internally accounted for by the LBWA, despite the fact that we later assume that there is zero amplitude in this region. Because $\omega$ is relatively far above $\omega^{(p)}$, the decaying Bloch mode in this region attenuates rapidly (see \textsection\ref{above_cutoff_sec} for details). The rapid attenuation means that the effect of this mode on the solution is negligible.

In figures \ref{fig:approximation_comparison_weak}(b) and \ref{fig:approximation_comparison_strong}(b), which correspond to the case where $\omega$ is just above $\omega^{(j)}$ (i.e. $\omega$ is a peak of $E_R$), we observe that the agreement between the semi-analytic and LBWA solutions is significantly worse. As was the case in figure \ref{fig:approximation_comparison_weak}(a), the most notable difference occurs in the region $pW<x<(p+1)W$. However, the increased value of $\omega$ means that $p$ has shifted from $j+1$ to $j$. Since $\omega$ is now very close to $\omega^{(p)}$, the decaying Bloch mode in the region $pW<x<(p+1)W$ attenuates slowly. We hypothesize that this slow attenuation allows the decaying wave to interact with the barriers at $x=nW$ for $n>p$ in a non-negligible way, which ultimately affects the solution.

\begin{figure}
    \centering
    \includegraphics[width=\textwidth]{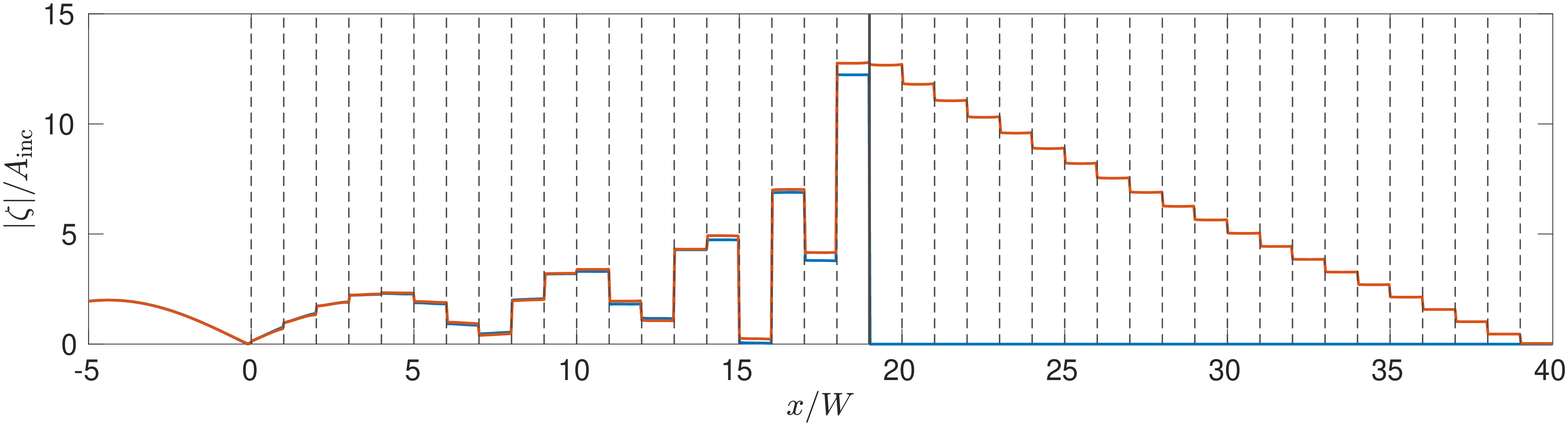}
    \caption{The free surface elevation in the graded array of vertical barriers described by \eqref{partial_grading}. As in figure \ref{fig:approximation_comparison_weak}, the elevation curves were computed using the semi-analytic method (red line) and our implementation of the LBWA (blue line). The positions of the vertical barriers are marked with dashed black lines and the position of the turning point is marked with a solid black line.}
    \label{fig:partially_graded_array}
\end{figure}

To support this hypothesis, we consider a partially graded array of vertical barriers. The partial grading is described by
\begin{equation}\label{partial_grading}
    d^{(n)}=\begin{cases}
        (n+1)\Delta d &\mbox{for }0\leq n< 20\\
        5\mbox{\,m}&\mbox{for }20\leq n\leq 40
    \end{cases}
\end{equation}
where $\Delta d=0.25$\,m. That is, the submergence depth of the barriers increases linearly up to $5$\,m, after which it remains constant. At $\omega=1.343$\,s$^{-1}$, the LBWA assumes that waves cannot propagate beyond the the $19$th barrier, since $\omega$ is in the stopband of this region. Importantly, this means that at $\omega$, our implementation of the LBWA is unable to distinguish between this partially graded array and the fully graded array with $\Delta d=0.25$\,m. On the other hand, the semi-analytic method can distinguish between these two cases. The free surface elevation in the partially graded array computed using both methods is given in figure \ref{fig:partially_graded_array}. We observe excellent agreement between the two methods to the left of the turning point, despite the LBWA failing to accurately predict the free surface elevation in the fully graded array at the same frequency. This highlights the importance of the part of the array to the right of the turning point when the decaying Bloch wave in $pW<x<(p+1)W$ attenuates slowly. Indeed, if the part of the array to the right of the turning point is not well approximated by a semi-infinite array (as the LBWA assumes), these decaying Bloch waves can be reflected into the region to the left of the turning point and affect the solution there. This suggests that the peaks of $E_R$ could be reduced by replacing \eqref{turning_point_reflection} with an equation that accounts for the grading of the barriers beyond the turning point, the derivation of which is not considered here.

\section{Concluding remarks}\label{discussion}
The LBWA has previously been established as a powerful tool for qualitatively understanding wave propagation through graded arrays. This paper has examined the LBWA in the context of graded arrays of vertical barriers by exploring the coupling mechanism between Bloch waves in adjacent unit cells. This mechanism was assumed to behave like the scattering of Bloch waves across the interface between two semi-infinite arrays of vertical barriers. We found that reflection by these interfaces is non-negligible when the differences between the submergence depths of the barriers of the two arrays is not small. This suggests that these reflections should be considered whenever the grading parameter of the metamaterial is significant. 

We then described a numerical implementation of the LBWA which accounts for these small reflections yet continues to omit decaying Bloch modes. The method accurately predicts the amplitude of the free surface at a wide range of frequencies. Exceptions occur in frequency regions which lie just above the cutoff frequencies of individual barriers in the array, where error peaks occur. We argued that these errors result from the LBWA failing to account for decaying Bloch waves in the region just beyond the turning point, which can have a non-negligible effect if they attenuate slowly. It may be possible to create a more accurate implementation of the LBWA by accounting for these attenuating waves.

We note that the large amplitude oscillations reported in this paper may not be realistic, since linear water wave theory is only accurate for low amplitude waves \citep{Mei2005} and the boundary conditions on the barriers are also greatly idealised when compared to real-world materials. The large responses in the array are related to the fundamental resonance of a pair of surface-piercing vertical barriers in isolation, which is known to become stronger as the ratio of the barrier submergence depth to the spacing increases \citep{newman1974interaction,mciver1985scattering,wilks2022rainbow}. Experimental or numerical tests (i.e. using wave flumes or computational fluid dynamics, respectively) would be required to evaluate the accuracy of the model.

The method presented in this paper may extend to graded arrays in which the scatterers are not vertical barriers, although further investigation will be required to confirm this. In particular, the LBWA could be implemented in the planar acoustics setting to study the coupling of (i) Rayleigh-Bloch waves in graded line arrays and (ii) Bloch waves in two-dimensional graded arrays. The concept of coupling two semi-infinite arrays discussed in this paper should generalise to these related problems, although they would be more mathematically complicated to solve.

\section*{Declaration of Interests}
The authors report no conflict of interest.

\appendix
\section{Bloch waves above the cutoff}\label{above_cutoff_sec}
The following discussion is heavily based on an article by \citet{bennetts2022rayleigh}, who studied Rayleigh-Bloch waves above the cutoff in a medium governed by the Helmholtz equation that contains an infinite line arrays of circular cylinders \citetext{also see \citealp{bennetts2017localisation,Bennetts2018,Bennetts2019}}. Our task is to show that in an infinite array of vertical barriers, the attenuation rate of Bloch waves above the cutoff is an increasing function of frequency.

\begin{figure}
    \centering
    \includegraphics[width=\textwidth]{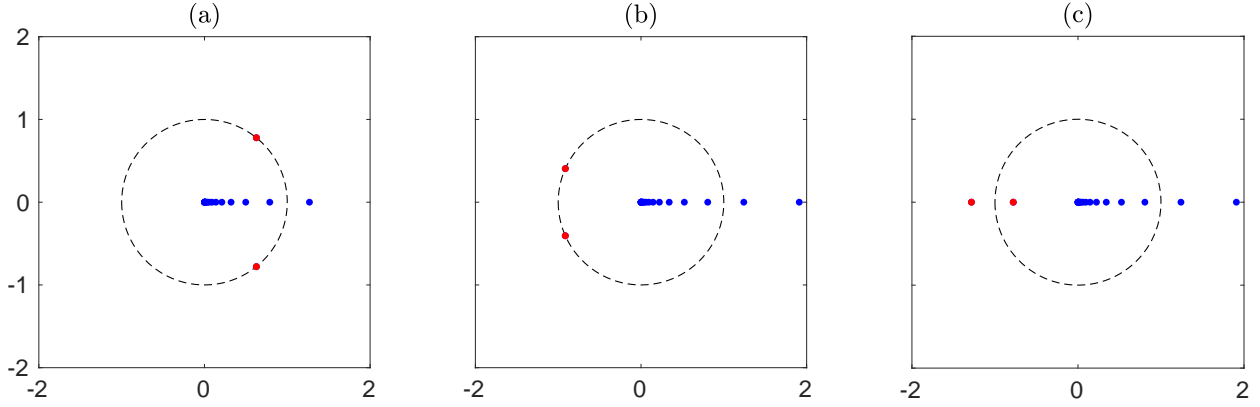}
    \caption{Generalised eigenvalues of \eqref{Bloch_matrix} for the infinite array of vertical barriers with $d=5$\,m and $W=2$\,m at (a) $\omega=1.2$\,s$^{-1}$, (b) $\omega=1.34$\,s$^{-1}$ and (a) $\omega=1.344$\,s$^{-1}$. The values $\exp(\pm \upi q W)$ are marked with red dots and all other eigenvalues are marked with blue dots. The unit circle is marked with a dashed black line.}
    \label{fig:eigenvalues}
\end{figure}

\begin{figure}
    \centering
    \includegraphics[width=\textwidth]{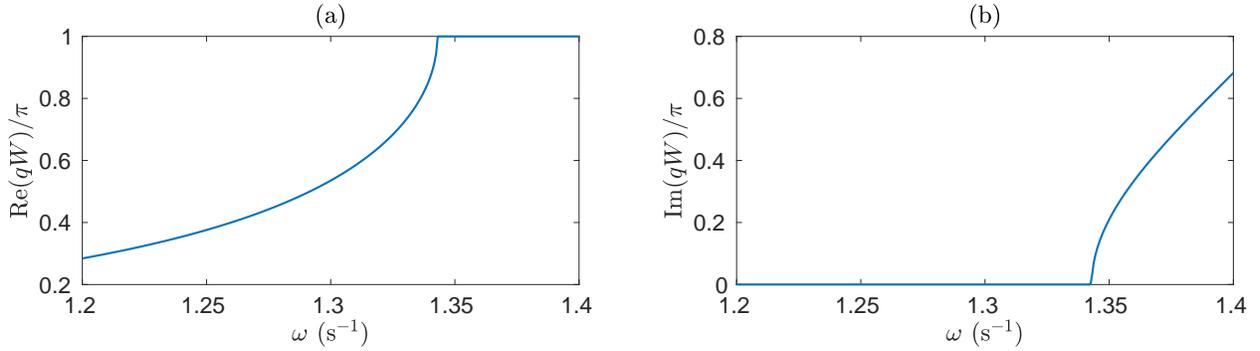}
    \caption{Plots showing (a) the real part of $q$ and (b) the imaginary part of $q$ as functions of the angular frequency.}
    \label{fig:attenuation}
\end{figure}

Recall that $\exp(\pm \upi q W)$ are eigenvalues of \eqref{Bloch_matrix}. Below the cutoff, $q$ is real and therefore these eigenvalues lie on the unit circle. The eigenvalue in the upper (lower) half-plane corresponds to the right-propagating (left-propagating) Bloch mode. As $\omega$ approaches the cutoff, $q$ approaches $\pi/W$ and both eigenvalues move around the unit circle, eventually merging at $-1$. Above the cutoff, the merged eigenvalues bifurcate as a real-valued reciprocal pair, with one moving inside the unit circle and the other moving outside the unit circle. The eigenmode corresponding to the eigenvalue on the inside (outside) of the unit circle describes a Bloch wave which decays towards the right (left). The decay rate of the Bloch wave above the cutoff is described by the imaginary part of $q$. In figure \ref{fig:attenuation}, we observe that above the cutoff, the imaginary part of $q$ is an increasing function of $\omega$, which is what we wanted to show. Although $d=5$\,m and $W=2$\,m have been fixed in figures \ref{fig:eigenvalues} and \ref{fig:attenuation}, we have verified that the result holds for a range of other values of the parameters.

\bibliographystyle{plainnat}
\bibliography{bibfile}

\end{document}